# “The Information as Absolute” Concept and Basic Physics

***Sergey V. Shevchenko[1] and Vladimir V. Tokarevsky[2]***
[1]*Institute of Physics of NAS of Ukraine, Kiev, Ukraine, ret.*
ORCID: https://orcid.org/0000-0003-3370-5536
[2]*Institute of Chernobyl Problems, Kyiv*

**Abstract** This paper is the presentation of the 2007-2026 Planck scale informational physical model, which is based on philosophical “The Information as Absolute” concept, which was formulated mainly in 2007. In the concept it is rigorously proven that nothing exists besides some informational patterns/systems of the patterns that are elements of the absolutely fundamental, and absolutely infinite, “Information” Set. Thus Matter for sure is nothing else than some informational system of informational patterns/sub-systems – particles, fields, bodies, etc. The other fundamental base of the model is the outstanding findings of von Weizsäcker and Fredkin-Toffli, who proved that Matter is based on some binary logics (“UR hypothesis”), and that if a system consists of reversible elements, then this system doesn’t dissipate energy outside, whereas the conception above makes these findings as completely natural. That allowed to define scientifically a number of fundamental phenomena/notions, first of all “Space”, “Time”, “Matter”, “Energy”, “Inertia”, and so to solve, or essentially to clarify, a number of fundamental physical problems, considering everything in Matter as some specific disturbances in Matter’s ultimate base – dense lattice of fundamental logical elements (FLE) that is placed in real Matter’s (here utmost universal “kinematical”) [5]4D spacetime) – what are particles and antiparticles, what are physical senses of basic equations in kinematics, first of all of Lorentz transformations and in dynamics, etc.

## 1. Introduction

In “The Information as Absolute” concept [1, 2, 3] it *is rigorously proven* that the phenomenon/notion “Information” (the rigorous definitions of phenomena “Information”, “Space”, and “Time” see below, here some instinctively used by humans, but adequate here enough definitions are used) is absolutely general and fundamental, when everything what exists is/are some “realizations” of some information – everything is/are some informational patterns/systems of the patterns that are elements of utmost general and fundamental absolutely infinite “Information” Set. The Set - i.e. all absolutely infinite “number” of its elements, exists always, “in absolutely long time”, having no Beginning and no End, since *it fundamentally – logically – cannot be non-existent*.

Every information of every the Set's element, i.e. every element itself, can exist on two main states – real and implicit ones, as concrete statement and the statement negation. An example the first is "this element really exists" and the second is "this element really doesn't exist", the first is true at concrete time moment, the second is true when relates to this element in time before, including before the element creation, and in future. At that both states contain practically identical information about element, though in compressed form, however absolutely completely. Say, "this element really exists" means that concretely this, this, this….sub-element in the element exists, interacts…, etc. i.e. the element, contains all true and complete information about element; the negation contains information that only partially different; and both are practically identical to informational pattern "element" before its creation as real.

Including in the case when, on first glance in the Set would not be anything - would be complete "nothing", and so the Set would be empty set, however that is fundamentally impossible – the Set logically inevitably contains the element – evidently true informational pattern "there is nothing"; which, though, is incomplete , more complete is the infinite cyclic statement "there is nothing besides the information that there is nothing besides…" - "Zero statement" in the conception. The statement in the compressed form contained, though only implicit, absolutely true, exact, and complete, information about all what will exist and happen in the Set in all absolutely long time, including true, exact, and complete, information about the real now Set's element "Matter".

The conception makes more clear a number of metaphysical and epistemological problems in science; first of all, the problem of the cognition - i.e. the problem of adequacy of the informational system "human's consciousness" inferences (in form of some language statements, including mathematical and algorithmic languages) to the reality - becomes be much more understandable, since the elements of the Set, including the system "Matter", which physics studies, are only some informational patterns/systems of the patterns also. Human's consciousness is fundamentally non-material, but she is a Set's element as well, and so there is nothing mystic in that humans study sometimes adequately to objective reality, including purely material, environment.

The Set and the phenomenon/notion "Information" have very unusual and interesting properties, including that the phenomenon and the Set are, in certain sense, the same, so both are here entitled. The informational systems/sub-Sets "Matter" and "Consciousness(es)" constitute the informational system/ sub-Set "our Universe".

In Matter the particles/objects, systems of objects/particles, exchange only by logically true informational "messages", and in accordance with a number of universal laws/links/constants, i.e. the informational system "Matter" looks something like as a computer. Such an idea isn't, of course, new - hypotheses that our Universe is a large computer appeared practically at once with the appearance of usual computers (see, e.g., [4 – 14], though the list can be much more), however that were only the hypotheses which had not necessary grounds (besides, of course, religious "hypotheses" of Creation of Universe as of a logical structure from nothing by some omnipotent sentient Being, Who "established the laws"). Including a number of papers that appeared last time, e.g., [15, 16], which again contain some seems as not too persuasive groundings only, as, e.g., [16] "...But now: what is the difference between Reality and its simulation? It's a matter for metaphysics: if Reality is indistinguishable from its simulation, then it is its simulation.

The Universe is really a huge quantum computer…" – such claims look as something more magic then scientific.

In the "The Information as Absolute" conception this idea becomes be grounded, moreover – the absence of logical structure of Matter (what realizes itself, including, as discovered by humans "Nature laws" in the Universe) would be rather surprising.

Note also, that really Matter isn't a "computer", it doesn't contain the computer's main functional modules – a processor, RAM, etc., which govern whole computer work; Matter is an "automaton" where every material object – particles, bodies, etc., "know" their functions in the system and behave at interactions, etc., with complete accordance with basic Matter's laws/links/constants, which so implicitly govern Matter existence and evolution.

Finally, in this section we give utmost common definition of the absolutely fundamental[1] phenomenon/notion "Information":

"**Information (data, evidences, signals, etc) is something that is constructed in accordance with the set/system of absolutely fundamental Rules, Possibilities, Quantities, etc. — the set/system "Logos" in the concept**".

Or, by other words, the "Logos" set elements "make something to be information".

A few examples of the "Logos" elements, which will be, since some of them do not have rational definitions in the mainstream science, scientifically defined and used further in this paper are, first of all, "Space", "Time", "Logical Rules", "Energy", "Change", and a few of others.

## 2. Physical model

### 2.1. A few common definitions

#### *2.1.1. Space and Time*

Space" and "Time" are absolutely fundamental Rules/Possibilities [elements of the "Logos" set] that are absolutely fundamentally necessary for any informational pattern/system could exist:

- "Space" is necessary for any information could exist at all, and

- "Time", additionally to Space, is necessary for some informational pattern/system could be dynamic, i.e. could change.

"***Space***" as the **Possibility** makes be possible placing in concrete "space" concrete informational patterns/systems, which (the space) at that is realized as a concrete set of "space dimensions", which (dimensions) *are necessary to actualize independent degrees of freedom of the concrete patterns/systems at changing of all their possible states*.

[1] Here and further the term "absolutely fundamental" relates to items that are valid/applicable in whole "Information" Set, the term ("only") "fundamental" relates to objects, phenomena, processes, etc., which are fundamental in the systems "Matter" and "consciousness".

Since Space is a logical possibility, the sets of dimensions form so concrete, and principally infinite, "empty space containers" for the concrete one type patterns/systems. For a space it is all the same – how many one type patterns/systems, which are constructed by the same concrete sets of logical rules/links/constants, and so have the same degrees of freedom at construction and changes, are placed in the container.

And it is all the same – in what places in the container the patterns/systems are placed. The unique requirement, when **Space** acts as the **Rule** is that a non-zero "space interval" must divide the different patterns/systems, and any pattern/system must occupy non-zero "space interval" [a "space volume", if there are more than one intervals in a number of different dimensions] as well. In that Space is the utmost universal grammar rule, which just so exists in all human languages.

Since any information absolutely fundamentally cannot be non-existent, everything had happened/existed in the "Information" Set; and everything is happening/existing, and will happen/exist always;

- the concrete patterns/systems, including Matter and consciousness, simply use the always existent concrete spatial dimensions from the absolutely infinite number of spatial dimensions of the Set's whole spacetime in concrete actualization of current state of concrete pattern/system. As that is, for example for Matter and humans in this concrete actualization of Universe evolution.

"***Time***" as the **Possibility** in main traits is analogue to Space, it is "the space for changing states of changing patterns/systems", and exists/acts in concrete cases forming, including, corresponding "time dimension" for dynamical patterns/systems.

However, Time has the essential difference from Space: for Time it is all the same for/ by what reason/way, by what degree of what freedom, etc., and in what informational pattern/system a change happened. So in this case it is enough to have *only one absolutely fundamental and universal dimension*, which exists and acts in whole "Information" Set for all changing states of all dynamic the Set's elements. In the concept and below the corresponding term is the "true time" dimension, since fundamentally erroneously used in conventional philosophy and sciences, including physics, "time" and "time dimension" really are specific space and space dimension, for this below term "conventional time" is used.

**Time** as the **Rule** also acts as that a non-zero "time interval" must be between different states of changing patterns/systems. However, in this case this Rule, unlike Space, seem as is determined by a couple of two, on first glance different, absolutely fundamental and "external to time" causes. The first one is that any information if appeared can not be non-existent, and so the next changing state can not "erase" previous state. The second is that a continuous changing of states is impossible, because of the logical self-inconsistence of the Change above, and the changes happen only along non-zero time intervals.

At any change of any informational pattern/system this pattern/system moves in the time dimension on corresponding time interval $\Delta t$, in every case, when the changing pattern/system is fixed in space, and at every change of its spatial position on, say, $\Delta x$. At that the changing of a pattern/system spatial position can be in principally arbitrary number of space dimensions, whereas all dynamic elements in the Set move at changes only in one, universal "true time" dimension (including in Matter's spacetime below).

A sequence of passed time intervals at changing states of the same pattern/system is motion of the pattern/system in the true time dimension.

Space and Time thus form concrete "empty containers" - "spacetimes", for concrete dynamical patterns/systems.

Finally, in this section we make a brief remark to existent definition of "Time" in recent physics. This definition was firstly done by Newton [17]

"…Absolute, true and mathematical time, of itself, and from its own nature flows equably without regard to anything external, and by another name is called duration …."

- at that for Newton, correspondingly, clocks show the time flow independently on time and only because of clocks tick equally equably,

- and this definition, however with the two relativistic modifications, remains in physics. According to special relativity postulates time (i) - not only flows equably, this flow depends on motion, and, whereas in stationary inertial reference frame time flows in accordance with Newton's definition, in moving frames its flow becomes be dilated, and (ii) – time governs material bodies, including clocks, and so "time is what clocks read", and clocks show in stationary frames "Newton's" flow, and in moving frames – dilated flow. Besides this time flow is observed as an "arrow of time" [18].

From the correct definition of "Time" above it follows that there cannot be any, "Newton's", "normal", "dilated", etc., time flows, and any specific "arrows of time" as well. Matter, and every material object/system, simply constantly, because of the energy conservation law, change, and so move in the true time, passing from a given states to mostly more probable states; when some changing is deterministic, that only connotes, that the probability is equal to 1.

Though from the definitions of "Space" and "Time" above it follows that the spacetimes of stable in the Set informational patterns/systems are absolute, in physics there exists the problem "is Matter's spacetime absolute or not"?

This problem did not exist in mechanics till the fundamental EM force was discovered, or even in first years after development of the Maxwell-Lorentz theory, where EM objects, events and processes existed and happened as some disturbances in some "ether", fixed in corresponding absolute Euclidian space. However, in late 1800s it became clear, that seems the application of very mighty relativity principle to EM processes and events results in some paradoxical consequences, as, say, the "relativity of simultaneity". It also seemed that because of the principle it is impossible really to observe absolute space and corresponding absolute motion of bodies.

H. Poincaré wrote about the absolute motion in "Science and hypothesis" [19]:

"… Again, it would be necessary to have an ether in order that so-called absolute movements should not be their displacements with respect to empty space, but with respect to something concrete. Will this ever be accomplished? I don't think so and I shall explain why; and yet, it is not absurd, for others have entertained this view… I think that such a hope is illusory; it was none the less interesting to show that a success of this kind would, in certain sense, open to us a new world…"

However, from that the absolute space even indeed cannot be observed evidently does not follow that it does not exist. Nonetheless that was postulated in the first version of the special relativity theory (SR) in 1905 [20]. It was also postulated that there is no corresponding ("luminiferous") ether, which would be placed in the absolute space, and be a base of some absolute reference frame. So the SR was – and is till now - based on one more postulate that all/every inertial reference frames are absolutely completely equivalent and legitimate.

From the last postulate any number of evidently meaningless physical, logical, biological, etc., consequences directly and unambiguously follow, the simplest one is the well known "Dingle objection to the SR" [21] and its more known and more complicated version "twin paradox" [22], etc. Correspondingly really seems in any pair of frames the frames aren't equivalent by some concrete way. Say, if a car crashed into a pole, that can be identically analyzed in both frames - where the pole is at rest and the car moves and where the car is at rest and the pole moves, but it is evident that the first frame is preferred one. Other example is universal –that all inertial frames aren't absolutely completely equivalent really was shown by Zeno yet 2500 years ago. Indeed, in all reference frames, where Achilles and the turtle move with different speeds, Achilles really leaves the turtle behind, in spite of that is logically prohibited, if the motion of both is continuous – well probably because of "illogical" $\Delta p \Delta x \geq \frac{\hbar}{2}$. But that is inessential in frames, where the turtle or Achilles are at rest; in such frames Achilles gets ahead of the tortoise without any logical problems.

Nonetheless because of action of really extremely mighty Galileo-Poincaré relativity principle all frames really are, though non-completely, but practically completely equivalent. So an observer in a frame principally doesn't know – moves his frame in space or not, and as a rule establishes that the frame's velocity is zero, and from the example above follows that such frame really by some way is preferred one besides the simplicity of its application. Though/ of course, the Dingle objection, etc., are fundamental, and there can exist only unique really completely preferred frames that are at rest in the absolute $3D_{XYZ}$ space, more see below.

From even one meaningless consequence, which directly and unambiguously follows from the SR postulates above, it completely rigorously follows by the "proof by contradiction" that Matter's spacetime is absolute; and that follows from the definitions of Space and Time in Sec. 2.1 above as well. However, these SR postulates have been, and are, stated as completely true postulates in physics till now.

Correspondingly observation of the absolute motion, i.e. the motion of a body in the absolute $3D_{XYZ}$ space, is only a technical task, which can be principally solved, as that is shown in this informational model, and the absolute velocity of a pair of clocks can be measured yet now [23], [24].

Finally note here, that the Set's elements in implicit form (say, before creation), and the real elements are some informational patterns systems, which so are made by all “Logos” set elements, including “Space” and “Time”, so when some element is created, it exists, interacts, etc. in absolutely long time already existent its space and changes in the absolutely universal in the Set time dimension. Including Matter was created in the already existent [seems yet partially] observed now space

*2.1.2. “Energy”*

Energy is the “Logos” set element [3], [25], which is absolutely fundamentally necessary for to change, including, of course, to create, of any/every informational pattern/system. That is because of the fundamental logical self-inconsistence of the other absolutely fundamental [also an element of the “Logos” set] phenomenon/notion “Change”:

- at every change of something its state is simultaneously former, recent, and future states, when all the states are different by definition. That is logical nonsense.

To overcome this logical prohibition of changes at every change it is necessary to pay by two points:

(i) – to change [including to create] some informational pattern/system it is necessary to spend some non-zero portion of “Energy”. However, that is not enough if the portion is finite, and so, besides,

(ii) – really at any change the changing state on some level/scale is uncertain – “illogical”.

From the above follows that in any dynamical system, including Matter, states of system’s elements always change discretely – are “quantized”, and just such objects, effects, processes, quantum mechanics studies.

Note, though, that the fact of impossibility of deterministic continuous changes of anything was proven more 2500 years ago by Zeno in his brilliant aporias, when Zeno, in fact, predicted the quantum mechanics.

Relating to QM, note also here, that the QM really ad hoc postulate, which is introduced aimed at fitting the theory with experiment, that “all given type particles are identical” in the concept obtains rational ground - this QM postulate is adequate to the reality because all such particles are copies of the same informational pattern, that is a typical situation in Information.

Energy remains be a mysterious element of “Logos” set that drastically differs from other “Logos” set ones, which rather evidently “make something be information” and look as rather understandable even at instinctive using of information. Say, “space” and “time” are main grammar rules that are in every humans’ language. Energy doesn’t make information, it only open by some mystic way already existent in the Set tincan “this Set’s element doesn’t exist really”.

However, that till now is not too essential in physics. The reason is that Matter is rather simple logical system, which is based on a limited set of fundamental and highly universal basic logical rules/laws, links, and constants (more see below), where the exchange by energy at material objects interactions is, in depth, highly standardized and universal, and the dependence of the action of Energy on difference of informational content in different material objects so is inessential, besides that there are, correspondingly, a few “forms of energy” – “kinetic”, “thermal”, “nuclear”, etc.

*2.1.3. Inertia*

Inertia is absolutely fundamental phenomenon that characterizes the logical resistance to changes because of the self-inconsistence of “Change” above. As energy, the inertia in Matter can be, and is, characterized; according to Newton, by the physical parameter “inertial mass”. Note here, that that has no relation to the existent in standard physics explanation of what is the inertial mass as some action of the Higgs field.

On an aside note a tenet, rather popular in official physics, that “energy and mass are two faces of one coin, one of them converts to another”. That is fundamentally incorrect. Both absolutely fundamental phenomena “Energy” and “Inertia” indeed co-exist always in every informational pattern/system, including in every material object, but they are fundamentally different, and so, say, at the interactions in Matter first of all energy transforms into/is distributed as energy, though with obligatory accompanying by transformation/distribution of inertial mass.

**2.2. Matter’s logical base**

Many authors [7], [8], [13-15], etc., point out that Matter in our Universe is some rather simple logical system (in the “Matter computer” rather simple program code runs). That follows from the fact that the number of Nature laws is not large, when laws themselves are rather simple and can be reduced to a number of groups of high-level symmetry; and that Matter’s base are some logical elements – “cellular automata”, etc., however all that were/are only some ad hoc premises, which have no rational base, and cannot be tested experimentally.

Rational premises can appear only in framework of the “The Information as Absolute” concept [1] - [3]: according to the concept Matter absolutely for sure is an informational – and so logical, system of informational also patterns and sub-systems, which are particles, fields, bodies, cosmological objects, etc.

The more concrete answer to what is Matter’s logical base with a large probability must be, and so is in this informational physical model, in accordance with two indeed fundamental findings in XX century:

- in accordance with the outstanding von Weizsäcker’s 1953-54 year “Ur-hypothesis” [26 - 28] that if Matter is based on fundamental depth on a binary logics, then the space should be 3D, and Matter’s spacetime indeed has 3 space *XYZ* dimensions. That was, on one hand, the outstanding hypothesis that explains why Matter’s space is 3D, and, on the other hand, the fact that the space is indeed 3D is the mighty evidence for that the hypothesis can be correct, and

- in accordance with the outstanding Fredkin-Toffli’s finding [5], who showed that if some patterns in a system are based on a reversible logic, the system changes at interactions in it without energy dissipation outside the system. In this case, Matter that would be dissipation somewhere in the Set; thus seems thrifty Matter’s Creator used this fact; and so in Matter the energy conservation law acts.

Correspondingly the concrete spacetime of the concrete binary informational system Matter has 3 “purely space” dimensions. Since this system is dynamical system, as that follows from experimental data, the spacetime has the “true time” dimension, *t*, which is absolutely universal and common for all dynamical elements of the Set. Further in this paper, for some reason (see below) instead of “*t*” for the true time dimension is mostly used “*ct*”, *c* is the standard speed of light.

Besides the dimensions above Matter’s spacetime has once more dimension, to implement the reverse sequences of changes, which are in a sense “non-legitimate” in the true time, as some “travels backward in time”, what is principally prohibited in the true time. The dimension is really a specific space dimension; however it is actualized in some traits (first of all it is invisible by humans really 3D sensors, as true time also) in the Matter like the true time. This dimension is called the “conventional time”, “$\tau$”, dimension in this informational physical model since that is just the “time what clocks show” [more see below], and mostly further for this dimension the metrics “$c\tau$” is used.

Thus the Matter’s spacetime is fundamentally absolute, fundamentally flat, fundamentally continuous, and fundamentally “Cartesian”, [5]4D spacetime with the metrics ($c\tau, X, Y, Z, ct$), where “$c\tau$” is the “conventional time” dimension, “*ct*” is the “true time” dimension, and *X, Y, Z* are 3 “ordinary” space dimensions – unlike the “conventional time” dimension, which really is a specific space dimension as well. The dimensions, as that is shown in Sec. 2.1.1 above, are principally infinite by definition of Space and Time.

*2.2.1. Actualization of the Matter’s logical base*

It seems quite rational to suggest that the dimensions of the spacetime relate to the degrees of freedom at changing states of some analogues of the von Weizsäcker’s “Urs”, though, of course, not literally: the [5]4D fundamental binary reversible logical elements (FLE), which compose placed in the absolute space above a [5]4D dense lattice of them.

Besides, in the model, basing on existent experimental data, it is postulated also that all the 4D FLE “sizes” (in the spacetime metrics above) are identical and equal to the Planck length, $l_P$, The changing of the binary FLE states “FLE flip time interval” is equal to the Planck time, $t_P$, Since neither space or time dimensions fundamentally haven’t some “inherent” measures, and so only relative (comparing with standards) measurements of space and time intervals is possible, if time dimension is *ct*-one – as that is in the model, the FLE size in time dimension is equal to Planck length, and in this case motion of material objects in the spacetime happens as “equal footing” in all 5 dimensions..

So in the model it is postulated, that at the fundamental elementary changing of state of some material object/system, i.e. at a the binary FLE “flip”, the information in Matter changes on one bit, and that is observed as changing of the physical action, *S*, and angular momentum of QM objects, on the elementary action/elementary angular momentum , $\hbar$ . Correspondingly here the question appears – so what is difference between physical action and angular momentum? , which in the model now isn’t answered, but seems fundamentally obligatorily should be answered at the model development in future.

This postulate of [5]4D FLE ether allows to postulate a few other really well scientifically rational assumptions.

### 2.3. Particles

In official physics, particles really are principally transcendent items – since they are some objects of the transcendent in conventional philosophy and sciences "Matter".

From the informational concept above and from experimental data that particles — which absolutely for sure are informational patterns/systems — are some objects that constantly change their states; however, at that, they are stable, it quite rationally follows, and in this model is postulated [1], that particles are some cyclic close-loop algorithms,

- which cyclically change their internal states with frequency ω[2] so that a particle has energy $E = \hbar\omega = mc^2, m$ is the inertial mass, $\hbar$ is the fundamental elementary physical action, reduced Planck constant, $c$ is the speed of light. This hypothesis appeared early in 1920 as the "the Zitterbewegung". de Broglie hypothesis [30, 31].

A few naturally suggested, and postulated in this model, rational conjectures follow from that above:

(i) – particles are some cyclic disturbances in the FLE lattice, which appear when a 4D momentum impacts on a lattice FLE, which, after the impact, "flips" further causing sequential flipping of neighbor FLE, etc.

To cause a flip – and the corresponding sequential flipping of ether FLEs along a straight 4D line is enough infinitesimal momentum when the "FLE flipping point" propagates in the 4D ether and 4D sub-spacetime with metrics ($c\tau,X,Y,Z$) with the speed of light, $c = l_P / t_P$. However, if the momentum, $P$, is not infinitesimal, the flipping point can not propagate with the speed faster than c.

Thus after non-zero momentum impact the FLE start to precess, and unidirectional "FLE flipping point" motion transforms into a "helical" one — of "FLE flipping point" of so some cyclic close-loop algorithm's – a "particle's" (Figures 1, 2 below) motion along some 4D "helix" of cyclic sequentially flipping – and so precessing, FLEs. Note also, that in this case the "flipping point" moves along "helix" with the speed $c\sqrt{2}$ , as the flipping of FLEs happens "diagonally", nonetheless the "helix front" moves along the impacting 4D momentum $\vec{P}$ direction with 4D speed of light, $\vec{c}$ .

However, some "a helix's 4D axis" seems (4D cross-product vector mathematically doesn't exist, however this is completely correct in mathematical "static" case, while the helix is dynamic) does not exist mathematically as a 4D vector in the 4D space, so the disturbance in the lattice propagates, possibly, as a bi-vector or a tensor. Nonetheless the propagation has the direction – the direction of the creating 4D momentum's vector. Besides, the "helix" of FLE lattice disturbance experimentally is observed as a pointlike particle interacting with other pointlike particles. It seems rational to suggest that "pointlike interactions" are interactions of the particles' FLEs, i.e. the "size of interaction point" is near Planck length, even though the disturbance "a particle" is not so pointlike, and the positions of the interactions points are randomly distributed in some non-pointlike spatial region.

[2] In earlier papers that is introduced as "informational currents"

Besides, the observed projection of 4D helical propagating of the FLE flipping point on the 3D space essentially determines that particles propagate in 3D space as "waves" (but interact as "points"); what is observed as the "wave-particle duality"

(ii) From the existing experimental data it seems rational to suggest (in first approximation, see point (i) above) that the "radius" of the "helix" is equal to the particle's Compton length $\lambda = \frac{\hbar}{mc}$, and the corresponding "helical" angular momentum of the particle's "FLE flipping point" is equal so to the Planck constant, $\hbar$ .

(iii) The always moving in 4D space and in parallel in true time, particles are, thus, some "gyroscopes" which are always oriented relating to the propagating direction, and

(iv) Note also, that it follows from the experimental data that there are two main types of particles in Matter, depending on the parental 4D momentums. In the model that are "S-particles", created by $3D_{XYZ}$ spatial momentums, and "T-particles", created by 4D momentums that have directed in the "conventional time", i.e. along the $c\tau$-axis, components.

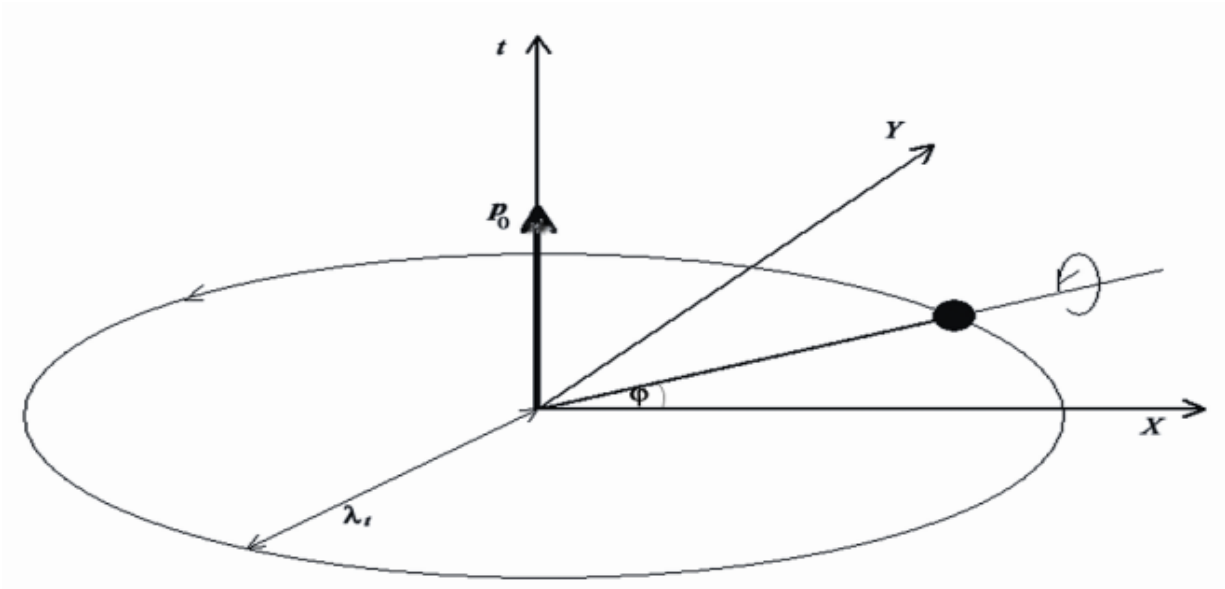

Figure 1. A T-particle at 3D space absolute rest. Large black point is flipping FLE on the circle that is the projection of helical movement of a particle along cτ-axis.

So S-particles, e.g., photons, always move in 3D space only with the speed of light, T-particles move in "conventional time" dimension with the speed of light, if are at rest in the absolute 3D space. If a T-particle moves also in 3D space after a space directed momentum impact, its speed in the "conventional time" dimension decreases by the Lorenz factor in accordance with the Pythagoras theorem.

Note, though, that the above in this section relates completely only to fundamental particles. If a particle is composed from some fundamental particles, some points in the above are are more complex.

And, besides, note that extreme impacts on FLE can result in many comparatively stable close-loop algorithms, and that is observed experimentally – the observed particles zoo now contains a more than a few hundred items – some chimeras that are composed from some fundamental particles, truncated particles' algorithms, as that, say, rather possibly muon and tau-lepton truncated electron's algorithms are; 2-nd and 3-rd generations of quarks, as well?, etc. Practically all of the algorithms have some defects, and so can break on some algorithm's tick with some probability, so such particles decay exponentially in time.

*2.3.1. Antiparticles*

In this model antiparticles are introduced by quite natural way – that are the same algorithms as the corresponding particles, however which run in reverse commands order. So the term "antiparticle" is really essentially applied only to T-particles/antiparticles, which are created by oppositely directed $c\tau$-components of 4D momentums, so algorithms run in opposite commands order directions, and antiparticles move along the $c\tau$-axis in opposite – n3gative - $c\tau$-directions. S-particles, say, photons, haven't different from the particles antiparticles.

*2.3.2. Particle's spin*

"Spin" is ad hoc introduced in QM as "purely QM "quantum number" - physical parameter of particles, aimed at fitting the theory with experiments. However, in the model it obtains rather "classical" sense – that is indeed an angular momentum, which is the observable projection of the "flipping point's" 4D angular momentum above on the 3D space. Thus, quite naturally spin can be – and really is in QM, added/subtracted to/from, say, "more classical" orbital angular momentum in atoms.

However, because of the mathematical limitation above, the 3D "angular momentum" "spin" observed in fundamental particles differs from the "real" the momentum's value, which is equal to $\hbar$, and for some T-particles it is observed at interactions in the 3D space as equal to $\frac{1}{2}\hbar$ ; fundamental T-particles are fermions. For S-particles the mathematical limitation above is not essential, and S-particles have the "real" spin, $\hbar$ (these are bosons). Though here is a limitation as well, the S-particles angular momentum can not have projection on the $c\tau$-axis, and so has only two $\pm\hbar$ spatial projections.

That the above relates only to the fundamental elementary particles, T- particles that are compositions of fundamental particles can have integer spins.

From the definition of what is the absolutely fundamental phenomenon "Inertia" above follows that all/every S and T particles have some inertia, and so all/every particles have inertial masses. But in this case there is a physically essential difference, though which is not principal: T-particles differ from S-particles in that they have inertial "rest masses", when S-particles quite naturally have not.

*2.3.3. Neutrinos rest masses*

From experimental data it follows [32] that the neutrinos are fundamental fermions, so are T-particles, and so have non-zero rest masses. Neutrinos, besides, since have extremely small masses, in real experiments move with speeds that practically equal to the speed of light, i.e. with large Lorentz factors. So their "flipping point" angular momentums, because of the rotation in the $(X, c\tau)$ plane (at motion along $X$-axis), are directed practically completely along spatial motion directions, and so are observed as be equal to $\hbar$. That is introduced in physics as that neutrinos have “helicity”. This helicity seem as practically for sure doesn’t differ from the helicity of, for example, electrons with large Lorenz factors, and, as that is for the electrons, here is no problem with reference frames – as all that must be in accordance with the relativity principle.

### 2.4. Lorentz transformations

If some T-particles constitute a rigid enough T-body (there are, though, no rigid bodies composed from S-particles), then, if the body is rigid, say, is a rigid rod, which has a length $L$, and is at rest in the absolute 3D space, the rod occupies a corresponding spatial interval equal to $L$, and all the rod’s points move in the $c\tau$-dimension with the speed of light, all the points so have identical $c\tau$ coordinate values.

However, if the body is impacted by some spatially directed momentum, as that always happens in mechanics, if we do not consider the interactions in high energy physics, then, as that was shown above, the rod’s speed in conventional time decreases in the Lorentz factor.

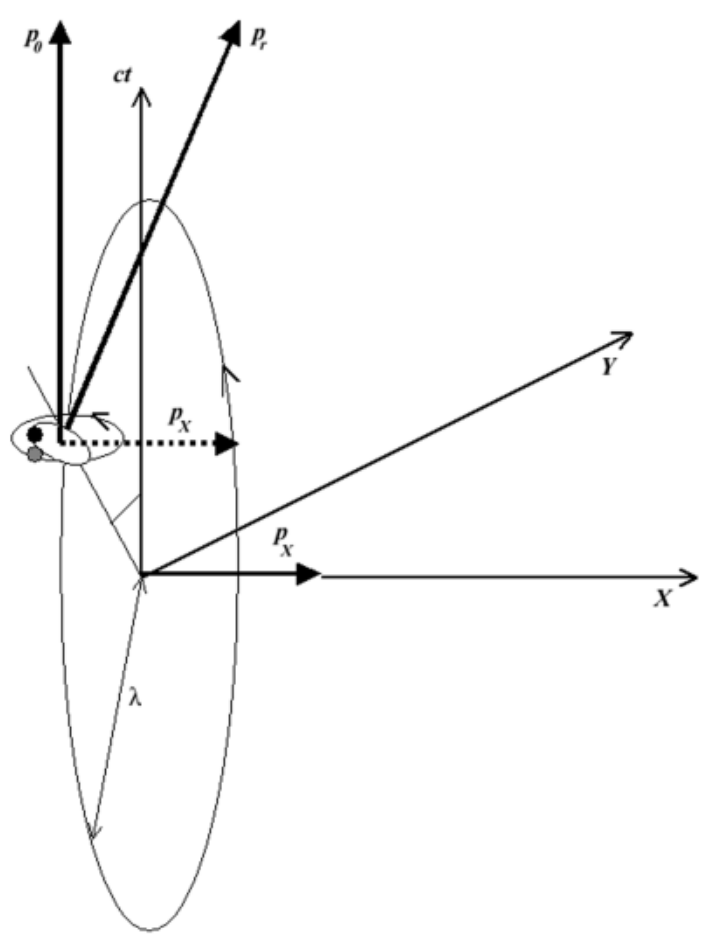

Figure 2. A T-particle’s (projection of helical) movement along $X$-axis as a combination of two circular and one direct motion. Big black points in the lesser circle are flipping T-FLE. Non – relativistic case. Really the motion isn’t some sum of independent helixes, that is rather bizarre “helix” of one particle’s algorithm, which is “diluted” by “blank” FLE flips in the “helix’s” large circle’s part.

Since the motion in conventional time is changing of internal state of particles (what is the running of the close-loop particles’ FLE-algorithms), the changing of internal state proceeds at maximal rate when a particle is at absolute rest, but when a particle moves also in the space, its algorithm is “diluted” by “blank” space steps, and so runs slower in the Lorentz factor.

Thus, the decrease of the moving particle’s speed along the $c\tau$-axis means that the internal processes in the particle are slowed down in the Lorentz factor as well. That is observed experimentally: moving unstable particles live longer, moving clocks tick slower, etc.; and, besides,

- since moving particles – "gyroscopes"– change their orientation in the 4D space the particles, if rigidly bound in the body, rotate the body in the space as a whole. So, in this case the rod, if it moves along *X*-axis with a speed *V*, rotates in the $(X, c\tau)$ plane on the angle, when the rod's front end becomes in the conventional time "younger" than the back end on rhe Voigt [34] decrement $-\frac{VL}{c^2}$ ("relativity of simultaneity", only along the rod and fundamentally nowhere in the space else), again in accordance with the Pythagoras theorem (Figure 3).

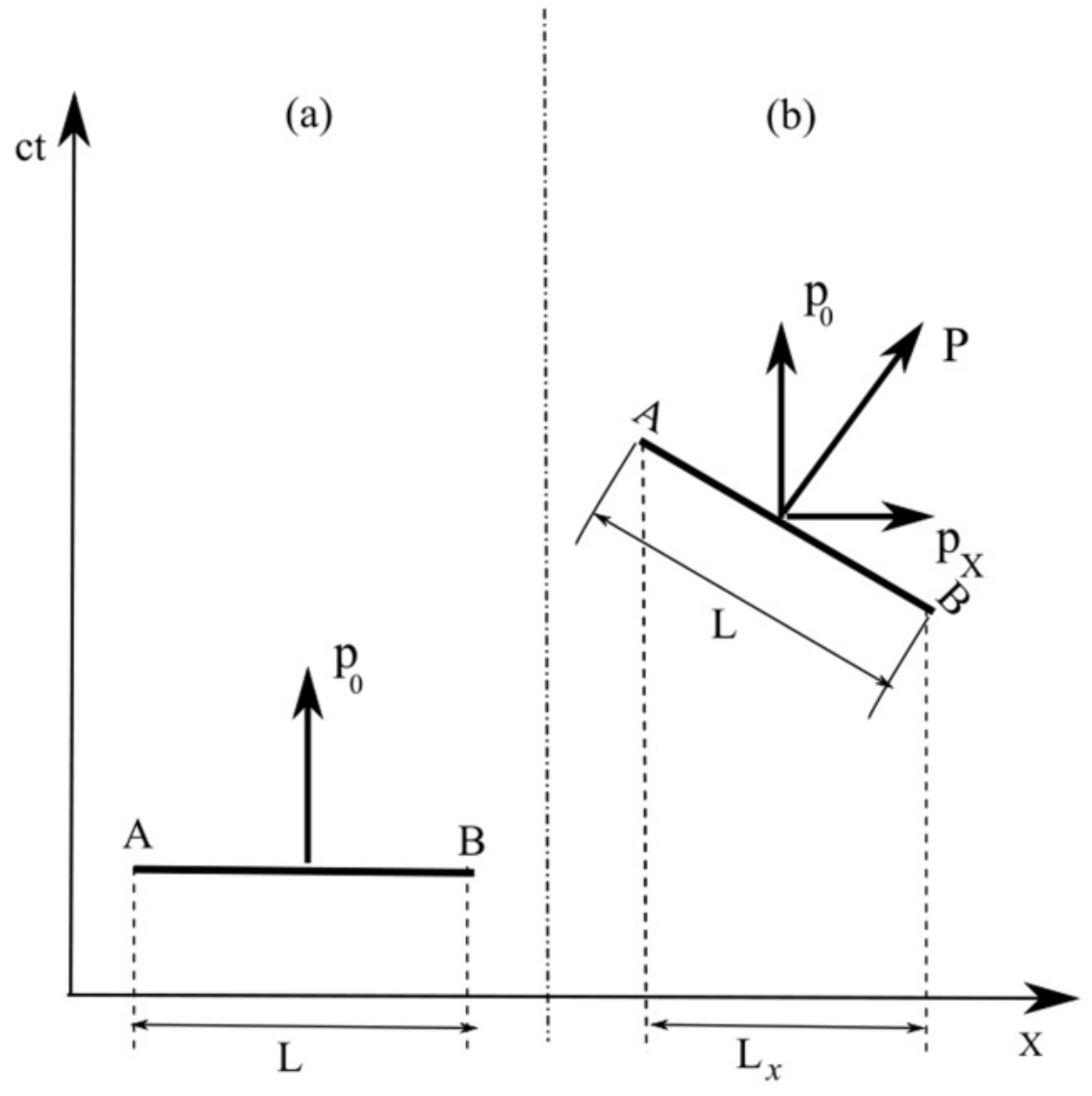

Figure 3. A rod having the length *L* and reat mass $m_0$ moves in the 4D space: (a) – the rod is at [3D spatial] rest (moves in the conventional time only) in the absolute reference frame, (b) the rod moves also along *X*-axis with a speed *V*. *ct* is the conventional time. $\boldsymbol{P_0 = m_0 c}$ is the rod's momentum when the rod is at absolute rest, ***P*** is the rod's momentum after the rod was impacted with transmission 3D space momentum $p_X$, , *m* is inertial mass of the rod. Internal processes in the rod, including the tick rates of clocks, located on the rod, are slowed in Lorentz factor

So the rod's projection on the *X*-axis is contracted by the Lorentz factor, as was suggested by FitzGerald yet in late 1800-th [34]. At that the rod really occupies a spatial interval $L_X$ in the space lesser than it occupied at rest, and all other material objects really interact with the "contracted" rod.

However, because the space interval etalons on the rod are contracted as well, the rod's length measured by an observer on the rod is again equal to *L*.

From the Figure 3 the Lorentz transformations [35], [36] immediately follow (note that till now all is considered in an absolute frame, where all parameters of bodies, in this case *x* and *t*, are real, and so, though clocks in the frame show principally the clocks' position in the conventional time, $c\tau$, clocks in absolute fame show also the position in true time, *ct*):

- the first equation

$$x = Vt + x'(1 - \frac{V^2}{c^2})^{1/2}, \qquad (1)$$

- and the second one:

$$ct' = (1 - \frac{V^2}{c^2})^{1/2} ct - \frac{V}{c} x' \qquad (2)$$

- which (the transformations) thus are indeed adequate to the reality. However, first of all:

- (in the above) the *Lorentz transformations are equations of motion of only points of moving in the absolute space rigid bodies in the absolute reference frame with using data about* coordinates *of these points, measured in the inertial reference frame* that is set on *(is co-moving) this body - as that in the Galileo transformations are*

The letters "$x/x'$", "$y/y'$", "$z/z'$", and "$ct/ct'$" in the transformations by no means relate to all points in the whole Matter's spacetime, as that is postulated in the Minkowski version of the SRT [22], and to points of some "local space" and "local time" in the Lorentz-Poincaré theory [35]- [37]; besides, of course, the spacetime points that are occupied by the bodies points.

So, of course, there is no "space contraction" – to affect on space is necessary to affect by some way on FLEs degreases of freedom, what for everything in matter is impossible fundamentally, there is no "time dilation" – to affect on true time for everything in "Information" Set, including, of course in Matter, is impossible absolutely. Though un this case really that relates to the "conventional time", but really it is the space dimensions, so see above in this passage; and really other "relativistic properties of the space, time, and spacetime" and corresponding "relativistic effects" don't exist.

Note, however, a few additional points in this case:

- first of all, because of the mighty Galileo-Poincaré relativity principle, which exists and acts because of the fundamentally binary reversible logical base of Matter, the Lorentz transformations form the group so that they are applicable not only in an absolute frame, they are symmetrically applicable between any the "Einsteinian" reference frames, i.e. that Einstein quite correctly used in first version of SRT in the 1905 year paper

"…The theory to be developed is based like all electrodynamics on the kinematics of the rigid body, since the assertions of any such theory have to do with the relationships between rigid bodies (systems of co-ordinates), clocks, and electromagnetic processes…"

- though after Minkowski illusorily postulated the applicability of the transformations to all the spacetime points, Einstein seems did not support further this 1905-year assertion about rigid bodies and frames' coordinate systems, and now the standard version of SRT is the Minkowski version [39].

Besides return to the SRT "complete equivalence of all inertial reference frames" postulate, which can have a sense only provided that there is no absolute Matter's spacetime; any number of senseless consequences logically rigorously follow from which. When we say about "[including inertial] reference frames" it is necessary to keep in mind that the frames are purely subjective humans construvtions, while for Matter it is all the same what humans think and do. Everything in Matter exists and happens fundamentally only in the absolute spacetime, where all parameters of moving, interacting, etc. material objects have real values, abd if the parameters would be measured in an absolute frame, clocks and rulers of which are at at rest in the $3D_{XYZ}$ space, the measured values of space and time intervals, so velocities, energies, etc. values would be identical with the real ones.

Including, say, if a rigid rod that if is at rest in the space has length $L$ is accelerated up to a speed $V$, measured in the absolute frame its length is lesser than $L$ in Lorentz factor, if clock moves st speed $V$, it ticks slower in Lorentz factor as well . At that if the observer in the co-moving wuth the rod frame measures its length, the measured value is $L$; and, if the observer measures length of identical rod that is at rest in the space, it obtains that the rod is contracted in the Lorentz factor, measured clock at rest tick rate is lesser than the frame's clock tick rate in the Lorentz factor. All the measured values in the moving frame are, of course, unreal, however, if observers in absolute and -moving frames will study what happens in same physical system using the frames instruments, the results of the researches will be identical, including so completely adequate to the objective reality.

At that , in each of the reference frames, the same clock ticks simultaneously faster (in this frame) and slower (in another), and if we assume that the systems are absolutely equivalent, as is postulated in SRT, this fact is obvious nonsense (this is the "Dingle objection to SRT") – but it is obtained experimentally,

Correspondingly, it should be remembered that "experimental confirmations" should be treated with a caution. However, we should note here that nothing fundamentally absurd occurs in Matter, and that what is above pertains to the operation of subjective products of human activity; and, furthermore, in this case, the observer in his "stationary" frame of reference clearly sees that when the observer in the moving reference frame measures the one-way speed of light, for example, along the direction of motion of his framem he is actually measuring velocities ($c\pm V$), and obtains the result $c$ only because (as he also sees and measures) in the moving frame, the rigid distances between the clocks are contracted, and the clocks tick more slowly, by the Lorentz factor, while the clocks located along the direction of motion showings are shifted by the Voigt-Lorentz decrement. The observer in another frame sees and measures the same things in the system above.

In this case, the ticking of the same clock simultaneously faster and slower is a real absurdity, and Dingle is absolutely right in this case, but the only real thing is that from his objection to the STR by proof by contradiction, the fundamental conclusion strictly follows - the spacetime of Matter is absolute and inertial reference systems are equivalent non-absolutely, there exist - as long as they can exist in principle - universally preferred absolute systems

However, in most cases this is unimportant in everyday physical practice, where reference frames are always rigid, and therefore the formalism of STR is quite adequately applicable - and at the same time does actually by no means this require the non-existence of absolute space-time.

It should also be noted that the concept of "rigid" in this case is not applicable literally, for example, the Earth's gravity makes the "Earth + satellites" system a rigid system, and therefore, for example, with the help of instruments in this system it is impossible to observe the absolute motion of this system.

If a system isn't rigid and is composed by free objects, the Lorentz transformations do not work completely. For example, that correctly is shown in the Bell paradox [40], where the space distance between free "Bell ships" doesn't want to contract. So by using such systems it is possible to observe the absolute motion, and to measure the absolute velocity, two corresponding experiments are proposed in [23], [24].

And, more importantly, the real non-complete adequacy of SRT postulates to the reality becomes an impediment in physics, when physics addresses the fundamental problems, i.e. outside the utilitarian applications in elaborations of concrete physical tasks and technology. Thus, new physics is possible in some cases only outside SRT.

A couple of examples, when really fundamental new results in physics turned out to be possible only as some violations of SRT, are the discoveries of antiparticles; and the "Feynman–Stueckelberg interpretation" in QED [40, 41], where it is postulated that antiparticles move backward in time, where

- Dirac's prediction of the antiparticles [42] is based on the suggestion that there are some points in "sea of negative energy", when "negative energy" does not exist in SRT (that does not exist at all, though),

- and moving of particles backward in time does not exist in SRT as well.

However, both these fundamental findings in physics remain unexplained, so really corresponding fundamental physical problems also remain. In spite of that the antiparticles predicted by Dirac are observed soon for 100 years already, they remain be some actualizations of rather questionable "sea of negative energy states" (in QFTs this sea now is as "false and true vacuums"); and the Feynman–Stueckelberg interpretation till now remains in physics as an ad hoc, really ungrounded, mathematical trick, which, however, is essential at application of very effective QED. Besides, cosmology, where 4D Minkowski space is postulated as real Matter's spacetime is based on really incompatible "main cosmological principle" and Big Bange hypothesis, which can be scientifically reconciled only provided that spacetime is [5]4D spacetime in this model above. Etc., more in this case see [25].

### 2.5. Dynamics

In sect. 2.3. it was shown that practically everything in Matter, i.e., particles, bodies, etc., is constantly moving in the Matter's "aether" – [5]4D FLE lattice with the 4D speeds of light, $\vec{c}$, having, correspondingly, 4D momentums $\vec{P} = (p_{c\tau}, p_X, p_Y, p_Z) \equiv (p_{c\tau}, \vec{p}_s)$, $p_{c\tau}$ is the $c\tau$-component , $\vec{p}_s$ is 3D space component of the momentum $\vec{P}$. As that is in Newton's mechanics, $\vec{P} \equiv m\vec{V} = m\vec{c}$ , *m* is the inertial mass.

So, because of that all dimensions in Matter's spacetime are fundamentally mutually orthogonal, $\vec{P} = \vec{p}_{c\tau} + \vec{p}_s$, where $\vec{p}_{c\tau} = mV_{c\tau}\vec{i}_{c\tau}$ is the $c\tau$-component, $\vec{p}_s = mV_s$ is the 3D space component of 4D momentum, $V_s$ is 3D spatial speed, $V_{c\tau}$ is the speed in the $c\tau$ - dimension, $\vec{i}_{c\tau}$ is the basic unit vector in $c\tau$-dimension, and $p_{c\tau} = m_0 c\vec{i}_{c\tau}$, $m_0$ is corresponding inertial mass, if the body is at absolute rest in the 3D space, "rest mass". Thus any impact on a body with only 3D space momentum – as that happens always in classical (outside high energy physics) mechanics – doesn't change the $\vec{p}_{c\tau}$ value, and so by Pythagoras throrem $P = (p_{c\tau}{}^2 + p_s{}^2)^{1/2}$, see Figure 4

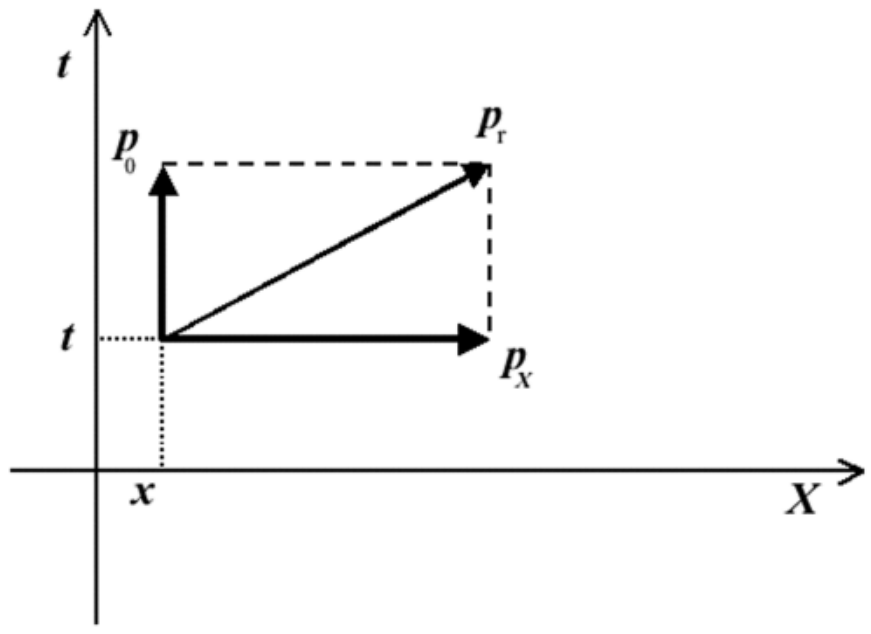

Figure 4. A momentum, $\vec{P}$, of a T-particle after space – directed impact with momentum $\vec{p}_X$. The particle moves along X-axis at a speed $V$

Correspondingly we obtain for the $\vec{P}$ magnitude another equation ($V \equiv \beta c$):

$$P = \frac{p_{c\tau}}{(1-\beta^2)^{1/2}}, \qquad (3)$$

and for $\vec{p}_X$:

$$p_X = P\beta = \frac{m_0 V}{(1-\beta^2)^{1/2}} \equiv \gamma m_0 V. \qquad (4)$$

where $\gamma = \dfrac{1}{(1-\beta^2)^{1/2}}$ is the Lorentz factor.

Calculating the work of some force $F$ at a spatial (a $c\tau$-direction impact results in the creation of new particles) acceleration of a body with a rest mass $m_0$ on a way $S$ (in the Equation (5) below $p \equiv p_X$ for convenience, in the point $S_1$ the force is equal to zero), obtain:

$$A = \int_{S_1}^{S_2} F(S)dS = \int_{p_0}^{p} \frac{p(1-\beta^2)^{1/2}}{m_0} dp = c\int_{p_0}^{p} \frac{p}{(p^2+m_0{}^2c^2)} dp = c\Delta P. \qquad (5)$$

Since at motion of a body the work of the force results in the change of the body's kinetic energy, from (5) we obtain

$$\Delta E = A = E - E_0 = cP - cp_0, \qquad (6)$$

or

$$E = cP = \frac{m_0 c^2}{(1-\beta^2)^{1/2}}, \qquad (7)$$

and for a body at rest in an ARF

$E_0 = cp_0 = m_0 c^2$ . (8)

Correspondingly in reality, the Hamiltonian in STR is an expression for the modulus of the momentum $\boldsymbol{P}$, multiplied by the $c$ constant.

The above is the derivation of the basic equations of dynamics. In addition, we note that from section 2.3 above it follows that every particle is some gyroscope, and so, for example, an impact on a particle (and on a body eventually) in some direction results in the particle's (body's) accelerations not along the impacting force direction, as that is in Newton's mechanics, but in two – along and orthogonal to the impact's direction – directions.

## 2.6. The informational model and "Euclidian relativity".

Many of the inferences above, obtained in the informational model, were presented also in a number of papers, where so called "Euclidian relativity" is developed [43-53]: "two face" nature of the time, the introducing of the absolute Euclidian spacetime and the absolute time ("Supertime" in [49]), etc. On another hand, in contrast to the model, the "Euclidian relativity" principles in the Refs above are introduced as some conversion of the SRT, as a rule by using the equation for the SRT invariant interval $ds$ :

$$ds^2 = (cdt)^2 - dr^2 \Rightarrow (cdt)^2 = ds^2 + dr^2 \Rightarrow (cdt)^2 = (cd\tau)^2 + dr^2,$$

where $t$ becomes be "Supertime" and $\tau$ - proper time – becomes be 4-th coordinate in Euclidian 4D spacetime. Such an approach seems as not totally rightful – the change " $ds \Rightarrow cd\tau$ " is valid for a material point only, in other cases proper time isn't applicable. Therefore, though the majority of the inferences of the approach in these papers are true, there are also others, for example in [41], [55] it is stated that relativistic equation for addition of velocities isn't correct. That isn't true; this equation follows from Lorentz transformations, which were obtained by Lorentz and further by Einstein for Euclidian spacetime, and just because of this equation for addition of velocities Lorentz transformations form the group, and so all inertial reference frames are mostly equivalent;

- and all that was before the introducing by Minkowski (what was found by Poincaré as interesting mathematical fact, that from invariance of this quadratic form ( $ds^2$ above) in a mathematical 4D space with imaginary time Lorentz transformations follow [38]) in the standard now SRT of the invariant interval and imaginary time.

Besides, what is more essential – when introducing the absolute spacetime the authors of "Euclidian relativity" apply, nonetheless, the principally erroneous relativistic approach, where the reference frame coordinates are something that directly corresponds to all/every points in the whole spacetime and so they introduce "4 D spacetime metrics", which depends on the reference frame; further – apply the coordinates rotations at transitions between, say, two frames at mutual relative motion. But in this case – analogously to the SR, where the rotations are principally inherent – it turns out to be that the reference frames simultaneously have two different (really $c\tau$ ) time-axes that have different directions – what is impossible in the reality, in the spacetime only one temporal axis exists. Thus in the reality really a number of the frames' coordinates transformations are possible – translations along any axis; and the rotations, however in 3D space only.

The next principal "relativistic" flaw, which is transmitted from SRT to the "Euclidian relativity" also – in both theories the coordinates of a reference frame (and the validity of Lorentz transformations that relates to the space and the time directly) are infinite, when in the reality the transformations (see above) relate only to corresponding kinematical parameters of motion of rigid material bodies and cannot be applied totally outside the bodies (see above section 2.4).

## 2.7. Informational approach and QM

Above we have noted an important, direct, and trivial consequence of the informational conception, which relates to the quantum mechanics – that one of basic QM postulates about the identity of the same type particles follows from that the particles, as everything in Matter, are some informational patterns, when any information has the property to have absolutely identical copies; and so one type particles are clones of one type algorithm.

Another basic QM point is the principal randomness/uncertainty of physical processes on a micro level.

That is the consequence of the fact, that the absolutely fundamental Logos set element "Change" (see section 2.1.2 "Energy") is logically self-inconsistent, and so this logical past-present-future problem" in dynamic informational patterns/systems, including in Matter, is solved by using energy, and by that dynamic processes on QM scale in Matter, are logically inevitably random/uncertain – "illogical.

However, this uncertainty isn't arbitrary. As is pointed above, Matter's objects change their states basing on binary logics, i.e., "bit by bit". From existent experimental data it seems to follow – at least till now - that there are no any experimental data inconsistent with this conjecture. In this informational model it is quite rationally conjectured, that on utmost fundamental depth all changes proceed as sequences of elementary steps on the Planck scale, when the physical action, $S$, is the number of binary operations, and every operation changes the information in a material object/system by one bit, observed as the change of the action on fundamental universal elementary physical action $\hbar$.

The Heisenberg inequalities in QM mean, essentially just that: $\Delta S = \Delta P \Delta x \geq \hbar / 2$ ,$\Delta S = \Delta E \Delta t \geq \hbar / 2$, etc.,

- however it looks as rather probable here is a correction: the inequalities seem with rather large probability really are the equalities.

So, though the QM uncertainty is absolutely fundamental, this uncertainty, nonetheless, is not arbitrarily random, and is actualized as a correlated uncertainties in pairs of non-commutative variables provided $\Delta S = \hbar$ in all cases.

Thus the "minimal physical action" principle in macro physics is that the states of interacting bodies proceed to change provided the minimal number of innumerous elementary binary steps with $\Delta S = \hbar$ (at every step information in the bodies changes on 1 bit); and by such a way QM directly reveals itself in macro physics.

Note also, that in this case it is necessary to take into account that this point is applicable in concrete informational systems, when some new informational structures, which have relatively self-dependent organization, appear. For example – elementary particles are some structures of FLEs that have properties, which the FLEs don't have (or, if be more correct – which FLEs have only implicitly, potentially); next level of the organization – atoms and nuclei that are some structures of the particles having new properties, etc., and on each next level new structures having new properties again must solve the "Change self-inconsistence problem".

**Wave-particle duality**

From the "helical" motion of particles, the experimental fact that particles move essentially as "waves" looks as quite natural, and so, for example, the diffraction patterns, if a beam of particles passes through an one slit, look as rather understandable. As well that experimentally particles interact as "points" is clear also now – at every time moments particles indeed exist as "FLE-flip points" - and interact as points that have size ~ Planck length, more see [25]

However an explanation of two and more slits diffraction possibly is essentially also outside this first approximation assumption, and will, rather probably, requite to assume also that a particle isn't a simple "one FLE line" algorithm, but at a motion the particle acts by some way on neighbour and further – other FLEs in the FLE lattice, when resulting "volumetric" disturbances of the lattice have properties that are inherent to the moving particle.

Finally mention an other QM puzzle – according to Schrodinger equation if a particle is at rest in 3D space, its Ψ-function isn't zero in whole infinite space? So the paricle can be in any oiunt in the space. That isn't of course, correct, any particle if is in rest in the space moves in $c\tau$ -dimension at speed of light, occupying in the 3D space region that has radius ~ its Compton length.More about other problems in QM, which this informational approach essentially clarifies see [25]

## 2.8. Mediation of the forces in complex systems

Now four "fundamental" kinds of the interactions (four "fundamental Nature forces") are experimentally known – Gravity, Weak, Electric, Nuclear/Strong; which differ by the strength, e.g., for the proton as (approximately) $10^{-36}:10^{-11}:1:10^{2}$. In conventional physics, the mediation of Fforces occurs through the exchange of energy and momentum via Forcea mediators. In classical physics, intense streams of mediators form Force fields that are continuously radiated by the charges of bodies (a concept dating back to the Liénard–Wiechert potentials of 1898–1902). In quantum field theories, these mediators are "virtual" particles—such as virtual photons in quantum electrodynamics, virtual $\pi$-mesons in nuclear physics, etc. —which are continuously radiated by the Forces charges of particles

. Nonetheless it looks as completely scientifically rational to suggest that in Matter there are no "virtual" particles and interactions, and the "virtual particles" really is a mathematical trick, which, for unknown now reason though, is – in QED extremely – effective at elaboration of physical tasks.

Other fundamental problem in both – classical macro physics and QM/QFTs is in that in the Forces theories it is postulated that the mediators and the fields contain/carry energy that is transmitted to "irradiated" particle/other charged body in the field, what looks as rather strange – to radiate constantly and always some energy particles/bodies must store some infinite energy.

Both the problems above are really scientifically rationally clarified in the initial basic Planck scale models of Gravity, Electric, Nuclear/Strong Forces, Weak Force looks as acts by some different way [25]. More see the this reference, here below some key points are.

Real interactions between particles in Matter indeed for sure happen at the exchange by the Forces mediators, which, however, are real and by no means "virtual". The Forces are some logical rules/marks, that can be, and are in Matter, assigned to, or, more correctly activated in, any FLE. If FLE is a "logical gate" in some particle's algorithm's FLE sequence, then at constant cyclic running of the algorithm, when this FLE flips, it causes flipping of neighbor the lattice FLEs which is specific in that, unlike to particles, the flipping propagates in the lattice not as a 4D helix, but as a 2D rim in $3D_{XYZ}$ space, in the rims FLEs corresponding Force mark is activated. thus the rims are the Force mediators.

The mediators are quite real, and, besides, since aren't particles, they don't carry energy. Note here also that this relates to corresponding, and additional to 4 utmost universal "kinematical" FLE degreases of freedom at changing its state above, so real Matter' spacetime is fundamentally absolute, fundamentally flat, and at least [4+4+1]8D Cartesian spacetime with the metrics ($c\tau$, $X$, $Y$, $Z$, $g$,$w$, $e$, $sn$, $ct$), where , $g$, $w$, $e$, $sn$, are Gravity, Weak, Electric and Nuclear/Strong, Forces dimensions.

Nonetheless the Forces govern by their "non-kinematical" mediators exchange, the exchange by kinematical momenta, and so by energies, between particles (so bodies…) at their interactions. More concretely this proceeds as follows:

- first, let us note that T-particles (i.e., those possessing a rest mass $m_0$) always have an energy $E$ equal to the sum of their "proper" (or "internal") energy $E_0=m_0c^2$ and their kinetic energy $E_k$; $E=E_0 + E_k = m_ic^2 = \gamma m_0c^2$, where $m_i$ is the inertial ("relativistic") mass and $\gamma$ is the Lorentz factor.

Force mediators are indeed constantly and continuously emitted by charged particles; however, since successive FLE flips in the rims propagate through the FLE lattice at the speed of light along straight radii, they do not possess—and thus do not transport—"kinematic" energy and momentum. Instead, they carry specific "non-kinematic" fundamental elementary momenta $\vec{p} = \frac{\hbar \vec{l}_P}{l_P^2}$, which, if a Force-tagged mediator collide with a flipping FLE belonging to the algorithm of an "irradiated" particle that has the same Force mark, changes in this particle its FLEs kinematical precession angle, so the particle algorithm changes its tick rate, so the particle's rest mass changes. That results in corresponding "actualization" in the particle of real kinematical momentum, and, if the particle isn't fixed, in the change of its $E_0$ and $m_0$, such that energy is transferred from $E_0$ to $E_k$ or vice versa. Since no additional energy is transferred, the particle's energy remains equal to $E$.

Let us consider two examples involving attractive forces (gravity, opposite electric charges, etc.) and repulsive forces (like charges, etc.); two particles with masses M0 and m0 (where $M_0 >> m_0$) interact while initially at rest in an absolute $3D_{XYZ}$ s[ace. The particles form a closed system and are initially separated by a nearly infinite distance.

Case 1: attraction. At a distance approaching infinity, two particles (or bodies…, respectively) are at rest in absolute $3D_{XYZ}$ space; their energies are $M_0c^2$ and $m_0c^2$, the total energy of the system is $E= M_0c^2 + m_0c^2$. After a very slight push towards M, particle m moves toward $M$; the direction of $m$'s motion coincides with the direction of the fundamental elementary momenta of the mediators, and with each collision between a mediator and a flipping the particle's marked by the attractive Force FLE, its internal energy—and thus its current rest mass $m_{0c}$ —decreases. At the same time, however, it acquires a **kinematic** $3D_{XYZ}$ momentum equal to the elementary momentum, and its kinetic energy $E_k$ increases; yet the particle's total energy $E = m_{0c}c^2 + Ek$ remains equal to the energy E the particle possessed at infinity. The same applies to M, except that its kinetic energy is much lower than the aforementioned $E_k$. Consequently, throughout all interactions, motions, and so forth within the system, its total energy remains constant—this is how the law of conservation of energy operates in Matter. For example, if a particle falls into a black hole, nothing strange happens to it— even when it crosses the Schwarzschild event horizon in GR; its energy remains equal to $m_0c^2$ (as measured at infinity), and the mass of the black hole increases by that $m_0$, even though the current $m_{0c}$ is much less than $m_0$ at infinity.

Case 2: repulsion. If the Force is repulsive, the particles do not form closed systems and can approach one another (reducing the distance between them) only if they possess appropriate non-zero kinematic momenta and velocities. As an example analogous to the one discussed above, consider the following case: a particle with rest mass m0, inertial mass $m_i=\gamma\, m_0$ , and kinetic energy $E_k$ moves from a distance approaching infinity toward a body of mass **M** (which is at rest in 3D space) with velocity $V$, such that $m_0c^2+ E_k = m_ic^2$. The energy of the system is $E_s = Mc^2 + m_0c^2 + E_k$.

The direction of motion of mass $m$ and the direction of the fundamental elementary momenta of the mediators are opposite; upon each collision between the mediator and the particle of mass m FLE , the particle's internal energy—and thus its current rest mass $m_{0c}$ —increases, while it simultaneously acquires a kinematic $3D_{XY}Z$ momentum equal to the elementary momentum and its kinetic energy $E_k$ and velocity decrease; however, the particle's total energy $E=m_{0c}c^2+ E_k$ remains equal to the energy the particle possessed at infinity.

When the particle comes to a halt in the field of $M$, its kinetic energy $E_k$ is zero current rest mass , $m_{0c}= m_i$ at infinity; so the energies of body $M$ and particle $m$ as well as the total energy of the system, remain equal to their values at infinity. When the particle moves back after stopping, the direction of its motion coincides with the direction of the fundamental elementary momenta of the mediators; consequently, the same process occurs as in Case 1—a transfer of energy from $m_{0c}c^2$ to $E_k$. At infinity, the particle possesses the same energy and speed $V$, albeit in the $V$ direction opposite to the original one.

Thus, the Forces primarily govern the exchange of momenta between the system's elements, resulting in a transfer of energy between the kinetic and internal energies of these elements. For example, at low speeds, the expression for the system's energy takes the form вид $E_s = p_M^2/2M + p_m^2/2m$ , $p_M = p_m$., where $p_M = p_m$. In other words, the energy transfer occurs "automatically" and is, in a sense, a process secondary to the exchange of momenta.

In static systems (such as atoms, planets moving in ideal circular orbits, etc.), essentially only an exchange of momentum takes place, while the energies of the system's interacting elements remain unchanged. However, such static systems—including those mentioned above and "fully static" ones—typically emerge following the evolution of open systems, a process that can involve (and often does involve) the dissipation of energy into the surrounding environment. Consequently, in such cases, the aforementioned equalities do not hold; the energies of the static systems (and their constituent elements) are lower than those of the "parent" systems by an amount equal to the binding energies of the elements.

To conclude this section, two observations are in order. First, we note that—as demonstrated by this Planck-scale model—the gravitational, electric, and nuclear (strong) Forces operate according to the same underlying mechanism; at the Planck scale, all interactions within matter are maximally unified, and the ratio of Forces strengths derived from the model matches experimentally measured values. Crucially, this specific ratio holds true only if the size and flip time of the FLE are equal to the Planck length and Planck time—as postulated in the model—thereby providing rigorous experimental validation for one of the model's most fundamental postulates.

Secondly, it should be noted that the above applies only to T-particles, which possess rest masses abd Force charges and emit Force mediators (or fields); in these cases, the particle energy remains constant during interactions (involving both particles and bodies) and does not exceed $m_0c^2$ at absolute rest. However, S-particles with zero rest mass—currently, this almost certainly refers only to the photon—lack charges (at least some do, such as the photon) and photons consequently do not emit Electric Force mediators. This is the case even though all the photon's algorithm FLEs likely bear the mark of this Force, and the photon itself is quite possibly created when, during the acceleration of an electric charge, certain mediators/ rim (in the "circular photons") receive "kinematic" momenta and transform into helixes. In all likelihood, the same process occurs with "circular gravitons" and "circular gluonons," although gravitons and gluonons have not yet been experimentally observed.

Accordingly, photons with energies lower than the e± pair production energy do not interact with “circular photons”, but interact directly with charged particles that have the E-marked FLEs in the algorithms without any restrictions, and, for example, thus accelerate particles in accelerators tubes up to TeV energies.

Besides , in certain cases photons can be applied "surgically"; for instance, if a proton “electrically” collides with another proton, it emits a stream of "circular photons" that interact simultaneously with all the quarks of the other proton, whereas photons can interact with individual quarks.

### 2.9. Planck mass particles

It seems worthwhile to mention here an additional remark, relating to the Beginning. There are, in principle, no objections to suggest [29, 32] that at the Beginning Matter was firstly created as a huge number of so called hypothetical “Planck mass T-particles”, algorithm of which has shortest logical length – 1 FLE, frequency $1/t_P$, rest mass energy $\hbar/t_P$, whole (“de Broglie”) algorithm logical length 6 FLEs (though, in principle closed sequence can be of 3 FLEs). So the particles have masses equal to the Planck mass ($m_P = \frac{\hbar}{l_P c} \approx 10^{19} BeV$). These particles contain (and their algorithm works by using) only the FLEs, which are absolutely symmetrical, and so these particles’ algorithms should be symmetrical also. Further iterations between/decays of these particles resulted in the appearance of observed now Matter. Such particles have at least two, possibly rather interesting, properties:

(i) – since for absolutely symmetrical algorithms it is impossible to choose a direction in the conventional time (“left” and “right” gyroscopes cannot be distinguished), *it is logically permissible* to assume, that *these T-particles at Beginning coild be creased by only positively directed in cτ-dimension momentums*, and.

(ii)– since these particles interact with anything only by extremely weak Gravity Force, there could be a part of the particles, which have not interacted at the Beginning (possibly ≈ 20% have interacted with the creation of observable Matter), and so now these primary particles can constitute, at least partially, so called “dark matter”.

Thus in such a case, if at Beginning only Planck mass particles (PM particles) were created, then in Matter there was no antimatter yet at Beginning, and further, at interactions and or decays of the primary particles under the angular momentum conservation law practically only some “ordinary” particles were created, whereas the part of antiparticles was rather small, and after corresponding annihilations and decay of unstable particles practically only stable particles remained – as that is at least in the observed now part of Matter.

Though, of course, the same situation occurs not only for the Planck mass particles, but if there would be some other particles with symmetrical algorithms; such case we cannot exclude now. However that practically nothing change in the consideration above, including in relation to the “dark matter” and the “matter-antimatter asymmetry” problems.

So it isn’t impossible that 70-80% of the primary matter exists till now in Matter’s space as PM particles with average density of the particles that is lesser then the density of baryons in $10^{18} - 10^{19}$ times. That is an extremely low density, so – and since Gravity is extremely weak force, and so the interactions cross-sections are extremely small – the probability of interactions of these particles now, say, in interstellar Space should be extremely small. Ordinary matter is highly transparent for the PM particles, they freely should pass through the bodies substance, rotating on individual orbits in halos around massive centers, which are formed by the ordinary matter, hqving increased flow density of PM-particles in perihelia of their orbits. Including so the DM particles really aren’t detectable, so rather numerous attempts to detect the particles experimentally failed, more see [55].

However, that could happen at Beginning, when after the “inflation epoch” in formed dense [5]4D FLE-lattice the energy was pumped in the lattice for creation of the primary particles, the energy was pumped globally uniformly, but non-uniformly locally, and corresponding compact local regions with enhanced primary particles density were some “seeds” of appeared further large cosmological objects; where, for example in centers of galaxies, in galaxies “supermassive black holes” these seeds, which contained only PM particles, till now constitute essential part of the SMBH matter; more see [25], section “Cosmology”.

## 3 Discussions and conclusion

Above a series of basic points in the initial Planck scale informational physical model is presented, which essentially differs from conventional physics, however this is rigorously scientifically substantiated. First of all – the model follows completely naturally from the philosophical “The Information as Absolute” conception [1, 3], whereas the truth, the completeness and the [self -] consistence of this conception are rigorously proven.

In the conception a number of fundamental phenomena/notions are rigorously scientifically defined – whereas really all such phenomena/notions in the conventional philosophy and sciences are fundamentally transcendent, and so essentially uncertain and irrational; including the important in physics, which studies fundamental phenomenon “Matter”, phenomena/notions “Space”, “Time”, “Energy”, “Change”, “Inertia”, “Matter” itself, “Information”, and, including because of known problems at interpretation of QM – “Consciousness”.

These phenomena/notions are introduced into physical theories as a matter of fundamental necessity; however, due to their transcendent nature, they are introduced by postulating their properties and effects. Such postulates are inevitably transcendent assertions — scientifically unfounded in a strict sense and fundamentally inadequate to reality—yet, because they are based on experimental data regarding the relatively simple logical system "Matter," they remain quite adequately applicable in scientific and technical practice. This is analogous to how the postulate that the Sun revolves around the Earth was applied adequately for over a century. Such was the state of physics until roughly the late 1980s; however, physics—specifically the Standard Model of particle physics and cosmology—has now begun to experimentally investigate Matter at a fundamental level where the transcendent nature of these postulates fundamentally impedes further scientific progress

All these phenomena in the conception cease to be transcendent **– in the concept it is completely rigorously proven, that absolutely for sure there exist nothing besides some informational patterns/systems of the patterns that are elements of the absolutely fundamental and absolutely infinite “Information” Set**. Everything is an information, whereas Information, in spite of is absolutely fundamental, isn’t transcendent, and any informational pattern/system so principally is study-able and cognizable.

Including Matter and human’s consciousness are some informational systems as well – both are made from only one stuff “Information”, and so, say, there is nothing surprising that some informational system, “human’s consciousness” is able to obtain some information about other, fundamentally different from consciousness, informational system, "Matter", and to analyze this information logically, in some cases finds in this informational system some logical laws/links/constants adequately to objective reality - despite that outside the concept all this humans do completely instinctively, really not understanding what is that they study, and what are they themselves.

At that Matter and Consciousness are fundamentally different dynamical informational systems, and so exist and operate in only partially intersecting spaces with mostly different dimensions, though evolve/develop changing their states in one common universal "true time" dimension. Albeit that consciousness evidently interacts with Matter, governing the practically material body, these interactions are weak, and so the consciousness really by no means impact on the material, including QM, objects/events/processes in physical experiments. If there is a bowl of sour cream in Schrödinger's cat box, everything that happens to the cat's wave function and the cat itself is completely independent of whether or not someone opens the box.

The conception completely rigorously scientifically rationally legitimates two really outstanding XX century Weizsäcker and Fredkin-Toffli outstanding findings, from which, and from the fact that existence and interactions of material objects – particles, bodies, cosmological objects– is determined in Matter by rather small set of basic laws/links/constants, it practically for sure follows, that Matter is based on some simple binary reversible logics, and, on some additional specific logical rules, which make Matter to be as it is – i.e. the composed from a few basic informational patterns, i.e. particles, nuclei/atoms, system of extremely diverse material objects, which [the rules] are observed, and called in physics "fundamental Nature forces",

More concretely from the outstanding findings quite rational assumption follows that the ultimately fundamental "material" base of Matter are (at keast) ) [5]4D binary reversible fundamental logical elements that compose the dense FLE lattice, which are placed in corresponding utmost universal (whole metrics is [4+4+1]4D) fundamentally absolute and fundamentally Cartesian, spacetime with [5]4D metrics *($c\tau$, X, Y, Z, ct)*, that particles are cyclic close-loop algorithms that run as constantly moving with the speed of light as FLE-by-FLE flipping in the $4D_{c\tau XYZ}$ space and in parallel in the common for all dynamic [and fixed, if are fixed at creation in some time point] "Information" Set's elements, ["true"] time dimension.

From this, and a few other concrete rational assumptions, it turns out to be possible to solve, or at least essentially, i.e. when possible ways of solutions are essentially determined, to clarify, a number of fundamental physical problems – besides scientific definitions of space, time, energy, inertia, which in conventional physics have transcendent fundamentally incorrect properties and originations:

- what are particles and antiparticles, including why neutrinos have rest masses; what is the physical sense of Lorentz transformations and special relativity theory at all; including what are really Lorentz factor, real and unreal, but observable, length contraction , slowing of internal processes rates in moving material objects, including moving clocks showings - and what are the "internal processes" at all; what is "relativity of simultaneity", etc.,

- In QM – why QM events/processes exist at all; why all given type particles are identical; what are "spin" and "wave-particle duality", etc. Besides quite rigorously scientifically rational initial Planck scale models of Gravity, Electric and Nuclear/Strong, Forces are developed, where gravitational, fine structure, and Nuclear/Strong force constants are derived; so – including, since in this whole Planck scale model main kinematics and dynamics equations and the fundamental experimental force laws of Coulomb and Newton- are derived, while the derived value of the nuclear interaction constant equals the experimental value – the model is in comp0lete accordance with all corresponding really reliable experimental data in physics; at that one of the main postulates in the model that the FLE size and flip time are Planck length and Planck time – and that FLEs compose the dense lattice, etc., though – is in whole accordance with experiment.

In the model also rather rigorously scientifically rational initial cosmological model of how Matter was created is proposed, what is fundamentally impossible in conventional physics. This model also is in full accordance with existent experimental data, and with all rational points in standard cosmology; including in the model the fundamental problem of matter-antimatter asymmetry is with well non-zero probability really scientifically clarified, more see [25], section "Cosmology".

According to the model Matter was created in the Set in a few steps, and the first step was the creation of only one (or a few) FLE, what happened in some point of the always existent whole [4+4+1}8D Matter's, by definition infinite, spacetime above.

Further rather probably, and that is postulated in the model, the process of programmed exponential expanding of the FLE lattice in $3D_{XYZ}$ space (and in parallel motion in *ct*-dimension , say, as sequential programmed division of FLEs on two ones – as that happens in bacterial colonies, had happened in extremely short time interval ("inflation epoch", in standard cosmology, however without, of course, some, "space expansion"), and rather probably all the lattice FLEs in the true time dimension turned out to be in one point.

Thus in the spacetime the dense FLE-lattice was formed, and after at the next step in the lattice the huge portion of energy was pimped, what resulted in creation of everything in Matter that is/are some disturbances in the lattice; which [disturbances] are always constantly changing, because of the energy conservation law, and so constantly moving simultaneously in the lattice and $4D_{c\tau XYZ}$ space with 4D speeds of light; and in parallel with the speed of light in the 1D time *ct*-dimension.

That formed the configuration of all material objects in the 4D space that evolved/evolves by the condition that every object from Beginning up to any concrete true time moment has passed the same way in the space and time, $S$, so that $S=\int_0^{t_{true}}|ds|=ct_{true}$, where $ds=(dx^2+dy^2+dz^2+c^2d\tau^2)^{1/2}$; the true time changes from the Beginning of Matter to given moment. That is a rather uncertain limitation, in every concrete case, since the interval for a concrete object – including, of course, the object's predecessors, can have any $[\pm]$ signs in all 4 dimensions at the object's evolution/transformation at its corresponding, sometimes bizarre, travel in the 4D space.

Nonetheless all FLEs in the disturbances are seems in one time point that has Planck length size, and so matter in Matter seems is in 4D surface of 5D sphere that has radius equal to *ct*. So the main cosmological principle, which is based on cosmological observations, that observed (really only in $3D_{XYZ}$ space) matter is distributed in Space uniformly and 4π isotropically, is incompatible with the Big Bang hypothesis, when older cosmological objects would be seen closer to the "B**ig Bang point**". I.e. it looks as that telescopes observe matter in postulated in cosmology 4D Minkowski space "as in biconcave lenses", while really adequate to the reality interpretations of the telescopes data are possible only in the [5]4D spacetime above.

Finely note that "the [5]4D – 4D spacetime" problem must be solved not only in cosmology, in both – classical mechanics and QM?QFTs – the universal Lagrangians and Hamiltonian techniques should be developed aimed at application in [5]4D spacetime and in accordance with that the Forces fields don't contain energy, etc. Though really fundamental problems utmost comprehensively will be solving only in next – after classic and quantum ones – Planck scale physics, development of which, with a well non-zero probability, will be based on this Planck scale model.